\documentclass[pra,reprint,superscriptaddress]{revtex4-1}
\pdfoutput=1

\usepackage{amsmath, amssymb, graphicx, hyperref, natbib}

\newcommand{\hb}{\hat{b}}

\newcommand{\+}{^\dagger}
\newcommand{\nodag}{^{\phantom\dagger}}
\newcommand{\hA}{\hat{A}}

\newcommand{\hH}{\hat{H}}
\newcommand{\ua}{\uparrow}
\newcommand{\da}{\downarrow}
\newcommand{\ket}[1]{\left| #1 \right\rangle}
\newcommand{\bra}[1]{\left\langle #1 \right|}

\begin{document}
\author{Bhuvanesh Sundar}
\email{bs499@cornell.edu}
\affiliation{Laboratory of Atomic and Solid State Physics, Cornell University, Ithaca, New York 14853, USA}
\affiliation{Department of Physics and Astronomy, Rice University, Houston Texas 77005, USA}
\affiliation{Rice Center for Quantum Materials, Rice University, Houston Texas 77005, USA}

\author{Todd C. Rutkowski}
\email{trutkow1@binghamton.edu}
\affiliation{Department of Physics, Applied Physics and Astronomy, Binghamton University, Vestal, New York 13850, USA}
\affiliation{Department of Physics, Radford University, Radford, Virginia 24142, USA}

\author{Erich J. Mueller}
\affiliation{Laboratory of Atomic and Solid State Physics, Cornell University, Ithaca, New York 14853, USA}

\author{Michael J. Lawler}
\affiliation{Laboratory of Atomic and Solid State Physics, Cornell University, Ithaca, New York 14853, USA}
\affiliation{Department of Physics, Applied Physics and Astronomy, Binghamton University, Vestal, New York 13850, USA}
\affiliation{Kavli Institute for Theoretical Physics, University of California Santa Barbara, Santa Barbara, California 93106-4030, USA}

\title{Quantum dimer models emerging from large-spin ultracold atoms}
\date{\today}
nearest-neighbor
\begin{abstract}
We propose an experimental protocol for using cold atoms to create and probe quantum dimer models, thereby exploring the Pauling-Anderson vision of a macroscopic collection of resonating bonds. This process can allow the study of exotic crystalline phases, fractionalization, topological spin liquids, and the relationship between resonating dimers and superconductivity subjects which have been challenging to address in solid-state experiments. Our key technical development is considering the action of an off-resonant photoassociation laser on large spin atoms localized at the sites of a deep optical lattice. The resulting superexchange interaction favors nearest-neighbor singlets. We derive an effective Hamiltonian in terms of these dimer degrees of freedom, finding that it is similar to well-known quantum dimer models, which boast a rich variety of valence bond crystal and spin liquid phases. We numerically study the ground state, explain how to tune the parameters, and develop a protocol to directly measure the dimers and their resonating patterns.
\end{abstract}

\maketitle
\section{Introduction}
Quantum dimer models---which describe the dynamics of close-packed hard-core dimers on a lattice---have received continued attention since their original proposal by Rokhsar and Kivelson in 1988 \cite{Rokhsar1988}. Several factors have motivated these studies, including connections to Pauling and Anderson resonating valence bonds \cite{Pauling1949,Anderson1973}, Anderson's theory of high-$T_c$ superconductivity \cite{Anderson1987,Lee2006}, the appearance of quantum critical points \cite{Sachdev1989,Senthil2004,Vishwanath2004}, topological order and fractionalized excitations \cite{Kivelson1987,Wen1991,Moessner2001a,Senthil2001,Fradkin2004}, their mapping to lattice gauge theories \cite{Moessner2000,Moessner2001b,Misguich2002}, and their potential applications in quantum computation \cite{Ioffe2002,Kitaev2003}. This rich variety of physics emerges due to the interplay between quantum fluctuations, hard-core constraints, and the lattice geometry of these systems. However, there are relatively few experimental realizations of dimer models. In this article, we show that experiments using an atomic gas trapped in an optical lattice can realize and probe dramatic dimer resonances in a range of quantum dimer models.

The spin physics emerging from atoms in optical lattices can be qualitatively different from those in electron systems. The atoms typically have spin greater than $1/2$, and the natural exchange processes lead to couplings which are more complicated than a simple Heisenberg model. As we demonstrate, these processes can be tuned to favor singlet dimers. 

Our technique works for both fermionic and bosonic atoms with vanishing electronic orbital angular momentum ($l=0$) and relatively large hyperfine spin $f$, but with relatively weak dipole-dipole interactions. Alkalis such as $^{7}$Li or $^{23}$Na are potential candidates. By tuning the lattice depth and trapping potential one drives the system into a Mott insulating state with one atom per site \cite{Jaksch2005}---effectively yielding immobile spins on each site which interact via a virtual superexchange process \cite{Anderson1950}. We propose manipulating these superexchange interactions by optically coupling pairs of atoms to an excited molecular state which has $L=1$ and $S=0$. When tuned sufficiently off-resonance, this optical coupling favors the formation of nearest-neighbor hyperfine spin singlets, which we refer to as dimers. In the large-$f$ limit the dimers are monogamous and orthogonal, e.g., a state where site $i$ forms a singlet with site $j$ is orthogonal to one in which $i$ forms a singlet with $k\neq j$. The resulting theory has the form of a quantum dimer model and, depending on lattice geometry and scattering length parameters, has the potential to realize dimer crystals and dimer liquid (resonating valence bond) ground states. At smaller $f$ the dimers are not orthogonal, but nonetheless the dimer configurations span the low-energy subspace. We show how to work with this nonorthogonal basis, and derive an effective dimer model.

We numerically find the ground state of our system for small lattices. When $f$ is large we find strong dimer crystal correlations indicative of the columnar dimer state on the square lattice and the $\sqrt{12}\times\sqrt{12}$ plaquette phase on the triangular lattice \cite{Ralko2005,Leung1996} (see Sec.~\ref{results}). At small $f$ the results are more ambiguous, and may point towards a spin liquid or a translationally invariant symmetry broken state (such as the nematic state predicted in Ref.~\cite{Rutkowski2016}). We explain how to further tune parameters to explore phase space---a useful requirement for the search for a spin liquid. We additionally propose a protocol to detect the dimer correlations which are central to many of these states.

To further characterize our model, we perform a large $f$ expansion, and find that as $f\to\infty$ it reduces to a special case of the Rokhsar and Kivelson model \cite{Rokhsar1988}. 
On the 2D square and 3D cubic lattice, there is some contention about the ground states of that model \cite{Banerjee2014,Sachdev1989,Leung1996}, an issue which an experimental realization of our proposal could resolve. 

There have been previous proposals to observe related physics in cold atom experiments, including crystallized dimer phases \cite{Demler2002,Yip2003,Imambekov2003,Zhou2006,Eckert2007}, resonating plaquette phases \cite{Chen2005,Xu2008,Szirmai2011}, and dimer liquid phases \cite{Hermele2009,Chan2012,Sinkovicz2013,Song2013,Rutkowski2016}---all hallmarks of the quantum dimer model. However, these studies were generally based on different mechanisms, and did not exploit mappings of their systems onto quantum dimer models. Additionally, Ising models can be implemented in cold atoms, and such models may be mapped onto dimer models \cite{kasteleyn}.

This paper is organized as follows. In Sec.~\ref{sec: microscopic model} we present the system we study, and its microscopic Hamiltonian. In Sec.~\ref{sec: OFR} we describe our proposal to tune the interactions. In Sec.~\ref{sec: model derivation} we present the effective model describing our system. 
In Sec.~\ref{results} we numerically find the eigenstates of this model, and describe their properties.
In Sec.~\ref{largef} we explore the large $f$ limit, mapping our system onto more traditional dimer models in Sec.~\ref{rkm},
showing how to tune parameters in Sec.~\ref{tune}, and describing the phases of this model in
Sec.~\ref{sec: phases}. In Sec.~\ref{sec: detection} we propose a method to observe these phases.

\section{Microscopic model} \label{sec: microscopic model}
\subsection{Setup}
To produce a quantum dimer model, we begin with a tight-binding Hamiltonian for atoms in an optical lattice, which includes spin-dependent interactions. There will be a hopping term, where an atom with hyperfine spin projection $m$ moves between neighboring sites. There will also be an on-site two-particle interaction term. In the presence of rotational symmetry, these interactions can be decomposed into different angular momenta channels $F$ \cite{Stamper-Kurn2013}. Thus, in complete generality we write
\begin{equation}
\hH = -J\sum_{\langle ij\rangle} \sum_{m=-f}^f\hb_{i,m}\+\hb_{j,m}\nodag + \sum_F U_F \sum_{M=-F}^F \hA_{ii}^{F,M\dagger}\hA_{ii}^{F,M},
\label{eqn: lattice}
\end{equation}
where $i$ runs over all lattice sites, and $\langle ij\rangle$ runs over all distinct nearest-neighbor pairs. Due to particle statistics, we sum over only even values of $F$, up to a maximum value $F=2f$ for bosons, and $F=2f-1$ for fermions. The $\hb^\dagger_{i,m}(\hb_{i,m})$ operators create (annihilate) an atom at lattice site $i$ with hyperfine spin $f$ and spin projection $m$, while the $\hA_{ij}^{F,M\dagger} (\hA_{ij}^{F,M})$ operators create (annihilate) a pair of atoms on sites $i$ and $j$ in total angular momentum state $F$ with total projection $M$. These operators may be defined via the relation
\begin{equation}\begin{array}{rcr}
\hA_{ij}^{FM\dagger} &= \frac{1}{\sqrt{2}}\sum_m C^{F,M}_{m,M-m}\hb_{i,m}\+\hb_{i,M-m}\+,& {\rm when}\ i=j,\\
 &= \sum_m C^{F,M}_{m,M-m}\hb_{i,m}\+\hb_{j,M-m}\+,& {\rm when}\ i\neq j.
\end{array}\label{eqn: Aij}\end{equation}
Here, $C^{F,M}_{m,m'} = C^{f+f\to F}_{mm^\prime}=\langle f,m;f,m' |F,M \rangle$ are Clebsch-Gordan coefficients, and the factor of $\sqrt{2}$ is chosen so that $\langle \hA_{ij}^{FM}\hA_{ij}^{FM\dagger} \rangle=1$ in the vacuum state. The kinetic energy term in Eq.~\eqref{eqn: lattice}---parameterized by the positive constant $J$---models the tunneling of atoms between neighboring lattice sites. The parameters $U_F$ encode the local spin-dependent interactions. While typically one expects that the scattering in different spin channels to be of similar magnitude, in Sec.~\ref{sec: OFR} we argue that one can engineer an optical Feshbach resonance so that the interactions are
significantly weaker in the hyperfine singlet channel than all others: $U_{F\neq0}\gg U_0>0.$ 

We refer readers to the review article by Stamper-Kurn and Ueda \cite{Stamper-Kurn2013}, for further background on Eq.~(\ref{eqn: lattice}).

\subsection{Effective nearest-neighbor inearest-neighbornteraction}\label{nn}
In the limit where the interactions are strong compared to the hopping ($U_F\gg J$), and there is exactly one particle per site, this system should form a Mott Insulator. Super-exchange will lead to a magnetic coupling between neighboring sites. In particular, we let $P$ be the projector into the space with one particle per site, and define $H_0=P H P$, $\Lambda=(1-P) H P$, $\Lambda^\dagger=P H (1-P)$, and $H_1=(1-P) H (1-P)$. Here $H_0=0$, and $\Lambda\propto J$ is considered small. We consider an eigenstate $\psi$, with $\psi_0=P \psi$, and $\phi=(1-P)\psi$. The Schrodinger equation $H \psi=E \psi$ can be projected into the space with one particle per site, and into the complementary space to give $H_0 \psi_0 + \Lambda^\dagger \phi = E \psi_0$ and $H_1 \phi + \Lambda \psi_0 = E\phi$. To lowest order in $J$, the second equation yields $\phi= -H_1^{-1} \Lambda \psi_0+{\cal O}(J)$, and hence $(H_0-\Lambda^\dagger H_1^{-1} \Lambda)\psi_0 = E \psi_0+{\cal O}(J^3)$, which yields the effective Hamiltonian $H_{\rm eff}= H_0-\Lambda^\dagger H^{-1} \Lambda$, or explicitly
\begin{eqnarray}\label{heff}
 \hH_{\rm eff} &=&
 \sum_F
 -\frac{2J^2}{U_F} \sum_{\langle ij\rangle} \hA_{ij}^{F\dagger}\hA_{ij}^{F}.
\end{eqnarray}
Under the condition $U_{F\neq 0} \gg U_0>0$, we can neglect all but the $F=0$ term to find
\begin{equation} \label{eqn: AdagA}
\hH_{\rm eff} \approx -\frac{2J^2}{U_0} \sum_{\langle ij\rangle} \hA_{ij}^{00\dagger}\hA_{ij}^{00}.
\end{equation}
On bipartite lattices this model is an example of a $SU(N)$ antiferromagnet model \cite{read1989some, harada2003neel}. We show in Sec.~\ref{sec: model derivation} that our model [Eq.~\eqref{eqn: AdagA}] can be mapped onto a dimer model.

For typical parameters (lattice depth $V_x=V_y=10E_R, V_z=30E_R$, wavelength $\lambda=1064\ {\rm nm}$, and scattering length $a_0 = 20\ {\rm Bohr}$), the superexchange coefficient is $\frac{2J^2}{U_0}=200\ {\rm Hz}$. This scale is large compared to neglected physics such as off-site dipole interactions ($\sim 0.5{\rm Hz}$ for alkali atoms). One may also worry about tensor light shifts from the lattice or photoassociation beams. These will be minimal if the laser detunings are larger than hyperfine splitting \cite{grimm2000optical}. If any residual light shifts remain, they can be canceled by adding additional fields (e.g., as in Ref.~\cite{gerbier2006resonant}).

\section{Tuning the interactions via an Optical Feshbach Resonance}\label{sec: OFR}
We propose inducing an optical Feshbach resonance \cite{fedichev1996influence, bohn1997prospects, ciurylo2005optical, nicholson2015optical, miller1993photoassociation, abraham1995photoassociative, abraham1996hyperfine, junker2008photoassociation, hamley2009photoassociation, gerton2001photoassociative, stwalley1999photoassociation, jelassi2006photoassociation, jelassi2006photoassociation2, kemmann2004near, wang1996photoassociative, wang2000ground, pichler2003photoassociation, fatemi2000observation, enomoto2008optical, yamazaki2010submicron, theis2004tuning, yan2013controlling, blatt2011measurement, kim2016measurements} between pairs of atoms by shining a laser tuned near a transition to an excited molecular state, labeled by orbital angular momentum $L=1$, electronic spin $S=0$, and total electronic angular momentum $J_a=1$. These are good quantum numbers in molecules formed from lighter elements such as Lithium or Sodium, where spin-orbit coupling is relatively weak [Hund's case (b)]~\cite{brown2003rotational}. For example, the laser can be tuned to couple the atoms to $^1\Sigma_{u/g}$ molecular states, as in Refs.~\cite{abraham1996hyperfine} and~\cite{fatemi2000observation}. The nuclear angular momentum is not important as long as the detuning of the laser is large compared to the hyperfine splitting. In the cold collision limit the rotational angular momentum of the nuclei vanishes, $R=0$. As in Ref. \cite{Sundar2016}, second-order perturbation theory then gives a contribution to $U_F = U_F^{\rm bg} + U_F^{\rm Fesh}$ of
\begin{equation}\label{eqn: UFFesh}
U_F^{\rm Fesh} = \alpha_F \frac{\Omega^2}{\delta+i\Gamma/2},
\end{equation}
with $U_F^{\rm bg}$ encoding the background scattering, including any influence of on-site dipole-dipole interactions.
The matrix element $\Omega^2$ is proportional to the intensity of the laser. The detuning $|\delta|$ must be taken much larger than the linewidth $\Gamma$, so molecular decay can be neglected \cite{hamley2009photoassociation}. This limit is reasonable, as in typical experiments $\delta \sim$GHz and $\Gamma\sim$100 MHz \cite{hamley2009photoassociation}. The line must be chosen judiciously so that unwanted transitions are avoided.

The coefficient $\alpha_F$ is the square of the overlap between the electronic spin singlet ($S=0$) and the hyperfine state with total spin $F$ \cite{Messiah1961}: It is related to the
Wigner $9j$ symbols. In Appendix~\ref{sec: deriving the effective interaction} we show how to calculate this matrix element for the most relevant case of alkali atoms, which have $s=1/2$ and nuclear spin $i$. We find
 \begin{equation}
 \alpha_F =\frac{(2i+1)(2f+1)-F(F+1)}{2(2i+1)^2}.
\end{equation}

Assuming that $U_F^{\rm bg}$ depends only weakly on $F$, one can then choose the laser intensity and detuning so that $U_{F\neq0}\gg U_0 \gg \left|\frac{\Gamma}{\delta}U_F^{\rm Fesh}\right|$, which then yields Eq.~\eqref{eqn: AdagA}. For Lithium, the laser intensity required to achieve this limit is only a few ${\rm W/cm}^2$.

\section{From Singlet Coverings to Dimer Models}\label{sec: model derivation}
\begin{figure}
\includegraphics[scale=0.27]{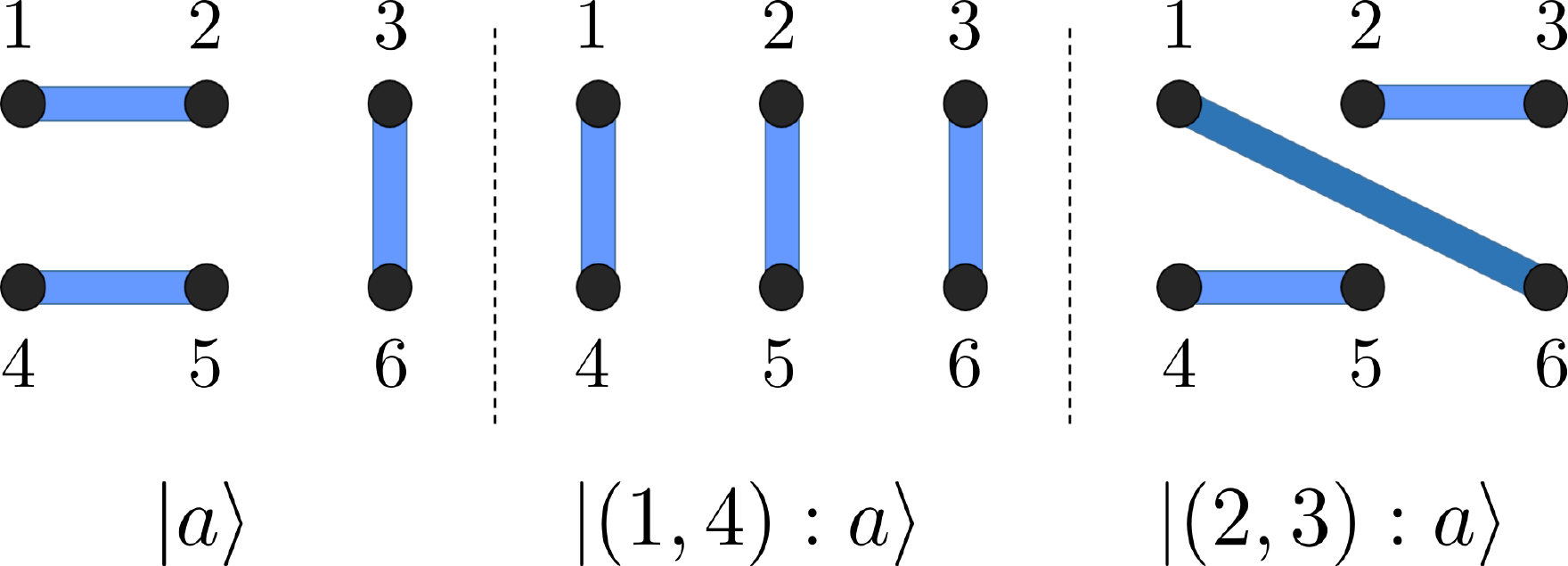}
\caption{\label{fig:ActionOnKets} (Color online) Examples of singlet cover states. The numbers label the lattice sites, while lines represent a spin singlet between the atoms on those sites. In this example, $\ket{a}=\hA\+_{1,2}\hA\+_{4,5}\hA\+_{3,6}\ket{0}$. The notation $\ket{(i,j): a}$, introduced in the main text, denotes a state where sites $i$ and $j$ are paired in a singlet, the original partners of $i$ and $j$ in $\ket{a}$ are paired in another singlet, and all the other bonds in $\ket{a}$ are left unchanged.
}
\end{figure}

The Hamiltonian in Eq.~(\ref{eqn: AdagA}) appears to count nearest-neighbor singlet bonds. One might therefore expect that the ground state would be formed by creating some pattern of nearest-neighbor singlets, which we will describe as a ``singlet covering.'' For example, the first image in Fig.~\ref{fig:ActionOnKets} illustrates one possible singlet covering of six sites that are laid out in a rectangle: 
$\ket{a}=\hA\+_{1,2}\hA\+_{4,5}\hA\+_{3,6}\ket{0}$, where $\ket{0}$ is the vacuum state with no particles. The label $a=\{(1,2),(4,5),(3,6)\}$ is the set of all bonds. In general,
\begin{equation}
 \ket{a} = \prod_{(i,j)\in a } \hA_{ij}\+ \ket{0}. \label{eqn: singlet covering}
\end{equation}
This definition works even for coverings which involve longer-range bonds (such as the third image in Fig.~\ref{fig:ActionOnKets}). The operator $\hA\+_{i,j}$ adds exactly one particle each to sites $i$ and $j$. We are working in the sector with exactly one particle per site, and therefore we require each site to appear in one and only one of the bonds. Furthermore, by the standard rules of adding angular momentum, a particle cannot be in a singlet bond with more than one other particle.

As explained by Rokhsar and Kivelson \cite{Rokhsar1988} in the context of spin-1/2 electrons, the singlet coverings are not eigenstates of the Hamiltonian, but they are closed under the action of Eq.~(\ref{eqn: AdagA}), and the ground state is a superposition of such coverings. In particular, a single term in the Hamiltonian maps one single covering into another:
\begin{align}
\label{eqn: Hamiltonian on kets}
\hA_{ij}\+\hA_{ij}\nodag\ket{a} &= \begin{cases}
 \,\,\ket{a}, & \text{for } (i,j)\in a\\
 \,\,(2f+1)^{-1}\ket{(i,j):a}, & \text{for } (i,j)\notin a ,\\
 \end{cases}
\end{align}
where the notation $\ket{(i,j):a}$ denotes a state where sites $i$ and $j$ are paired together into a singlet, the original partners of $i$ and $j$ in $\ket{a}$ are paired together into a singlet, and all the other bonds in $\ket{a}$ are left unchanged. An example of a singlet covering $\ket{a}$ and a few of the related states $\ket{(i,j):a}$ are illustrated in Fig.~\ref{fig:ActionOnKets}. Note, the labeling is not unique: One goes from $\ket{a}$ in Fig.~\ref{fig:ActionOnKets} to $\ket{(1,4):a}=\ket{(2,5):a}$ by either acting with $\hA_{14}\+\hA_{14}\nodag$ or $\hA_{25}\+\hA_{25}\nodag$.  As illustrated by the right-most figure, the nearest-neighbor bond operators acting on a state with nearest-neighbor bonds can generate configurations with longer-ranged bonds.

Thus, the Hamiltonian is a map on the space of singlet coverings.
Somewhat complicating the analysis, however, is the fact that the singlet coverings are not orthogonal. In fact, they are not even linearly independent. Nonetheless, as detailed below, it is straightforward to work with these states. 
In Appendix~\ref{sec: orthogonal basis} we review the more traditional approach of orthogonalizing the states, which is somewhat more involved. Working with either basis gives equivalent results.

We consider a general state $|\psi\rangle=\sum_a \psi_a |a\rangle$. 
Equation~(\ref{eqn: Hamiltonian on kets}) allows us to write
\begin{equation}\label{hmat}\hat H|a \rangle = \sum_{b} |b\rangle H_{ba}. \end{equation}
Clearly if $\sum_a H_{ba}\psi_a = E \psi_b$, then $\hat H|\psi\rangle=E|\psi\rangle$, and the eigenstates of the
matrix $H_{ba}$ either yield eigenstates of $\hat H$, or are null-vectors. The latter have eigenvalue 0. We will solely be concerned with states with negative energy, and hence will not encounter any of these null-states. Because of the nonorthogonality, $H_{ba}$ is not a Hermitian matrix---but it is self-adjoint with respect to the natural inner product: 
$\langle \phi|\psi\rangle = \sum_{ab} \phi_a^* \psi_b S_{ab}$, with $S_{ab} = \langle a|b\rangle$.

In Sec.~\ref{results}, we consider a small system, and enumerate all dimer coverings. Although exponentially large in the system size, this is a much smaller Hilbert space than a spin model on the same lattice. We then numerically calculate $H_{ab}$, and find its eigenstates.

If we formally set $f=\infty$, then the dimer coverings are orthogonal, and become eigenstates of the Hamiltonian. The energies of these states are negative, and proportional to the number of nearest-neighbor dimers. Thus the ground state manifold is highly degenerate, consisting of all nearest-neighbor coverings.
 In Sec.~\ref{largef} we derive a systematic expansion in $1/(2f+1)$, and find that the leading terms break this degeneracy and stabilize various dimer crystal or plaquette phases. The structure of this expansion corresponds to Rokhsar and Kivelson's model \cite{Rokhsar1988}.

In the limit of small $f$, we anticipate the dimer crystal order to disappear. In particular, Rutkowski and Lawler~\cite{Rutkowski2016}, carried out a variational study of the Hamiltonian in Eq.~\eqref{heff}, and argued that for $f<3$, a translationally invariant nematically ordered state will be found. Another, even more exciting possibility is a spin liquid---which could either occur as an intermediate phase, or at the phase boundary. 

Related physics is seen in studies of anisotropic 2D lattices of coupled spin-1 objects~\cite{harada2006, grover2007}. In those studies, the researchers finds regions with nematic order, and others with dimer crystal order. Surprisingly, there appears to be a direct second-order phase transition between these phases: Within the Landau paradigm such a direct transition would require fine-tuning. Moreover there is evidence that this transition displays ``deconfined quantum criticality,'' where the transition is described by an emergent gauge theory, and is a spin liquid~\cite{Senthil2004}. By analogy, one might expect that our model would display a similar critical point as $f$ is changed. We have, however, not yet verified this conjecture.

\section{Numerical Results}\label{results}
We numerically diagonalize the matrix $H_{ab}$ in Eq.~(\ref{hmat}). To visualize the ground state, we calculate the correlation functions 
$C_{ij}^{kl}=\langle \hA_{ij}\+\hA_{ij}\nodag\hA_{kl}\+\hA_{kl}\nodag\rangle$. As we argue in Sec.~\ref{sec: detection} this is an experimental observable. It corresponds to the probability of simultaneously having a dimer on bonds $(i,j)$ and $(k,l)$.
In a state $|a\rangle$ consisting of a single dimer covering, the expectation value $\langle a| \hA_{ij}\+\hA_{ij}\nodag|a\rangle$ is equal to $1$ if $(i,j)\in a$ and $1/(2f+1)^2$ otherwise. Similarly, $\langle a| \hA_{ij}\+\hA_{ij}\nodag\hA_{kl}\+\hA_{kl}\nodag|a\rangle$ is equal to $1$ if both $(i,j)\in a$ and $(k,l)\in a$, and is otherwise suppressed by factors of $1/(2f+1)^2$.

\newlength{\spc}
\settowidth{\spc}{(a) f=100}
\newlength{\spcb}
\settowidth{\spcb}{(b) f=3}
\begin{figure}[t]\centering
\includegraphics[width=1.\columnwidth]{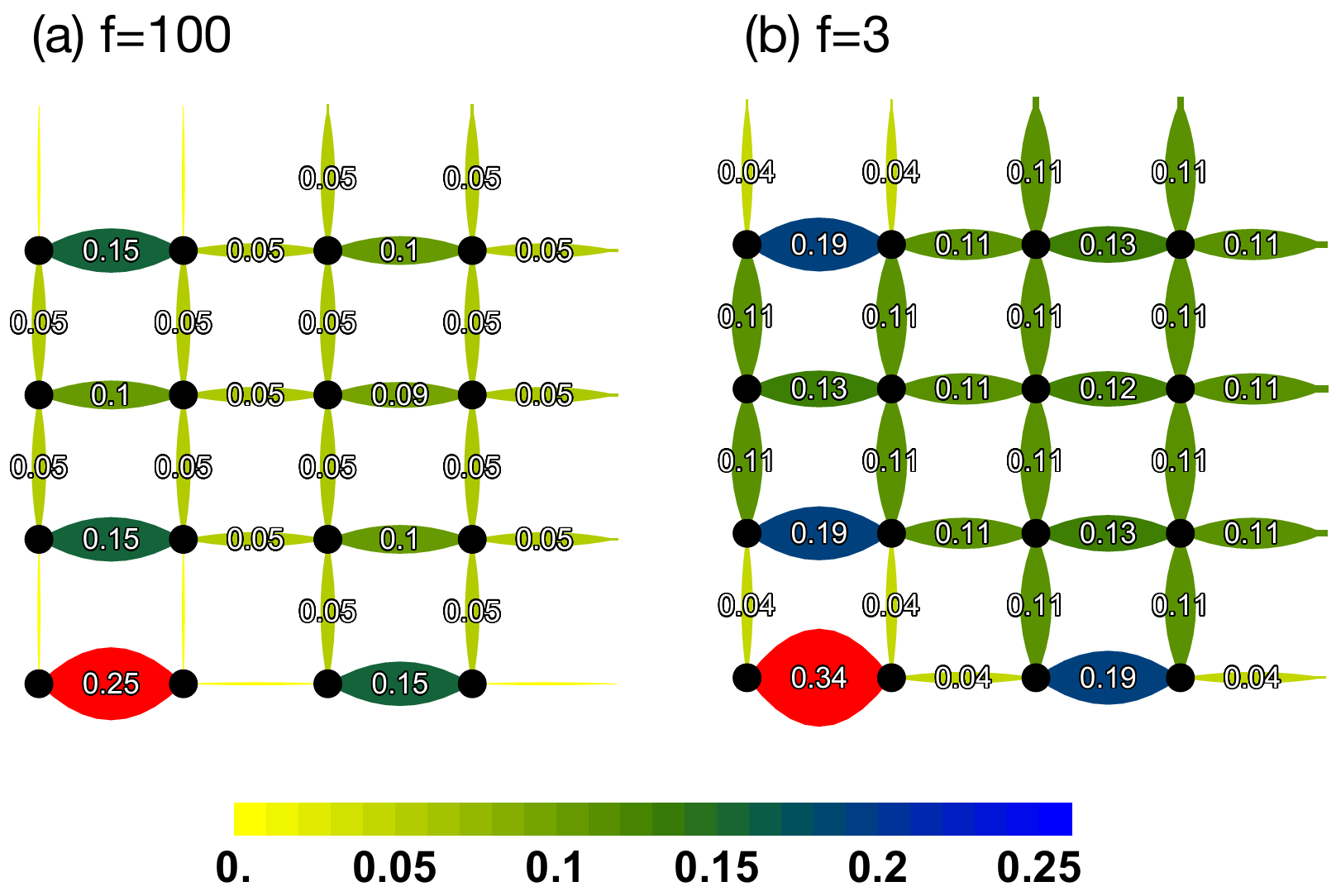}
\caption{(Color Online) Dimer-dimer correlations $\langle \hA_{ij}\+\hA_{ij}\hA_{kl}\+\hA_{kl}\rangle$ on a square lattice with periodic boundary conditions, for (a) $f=100$ and (b) $f=3$. The reference dimer $(i,j)$ is the fat red bond in the lower left corner. The thickness of the lines is proportional to the strength of correlation, which is also indicated by color.}
\label{fig:square lattice}
\end{figure}

\begin{figure}[t]\centering
\includegraphics[width=0.9\columnwidth]{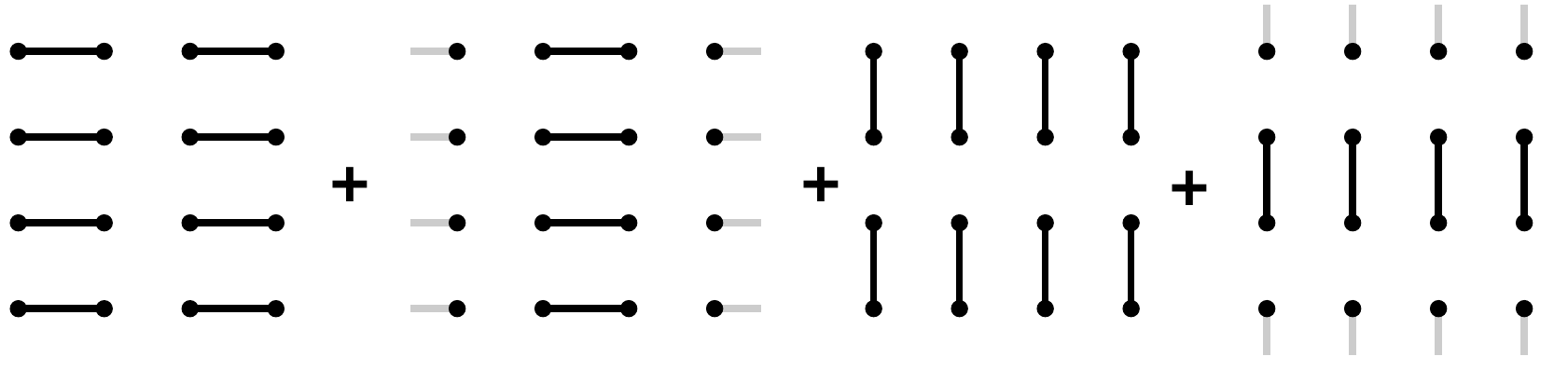}
\caption{Cartoon of the idealized columnar state, which is a superposition of four symmetry related ladder configurations. Light gray lines either represent bonds extending to sites beyond those shown, or to bonds that wrap around periodic boundaries. }\label{columnar}
\end{figure}

\begin{figure}[t]\centering
\includegraphics[width=1.\columnwidth]{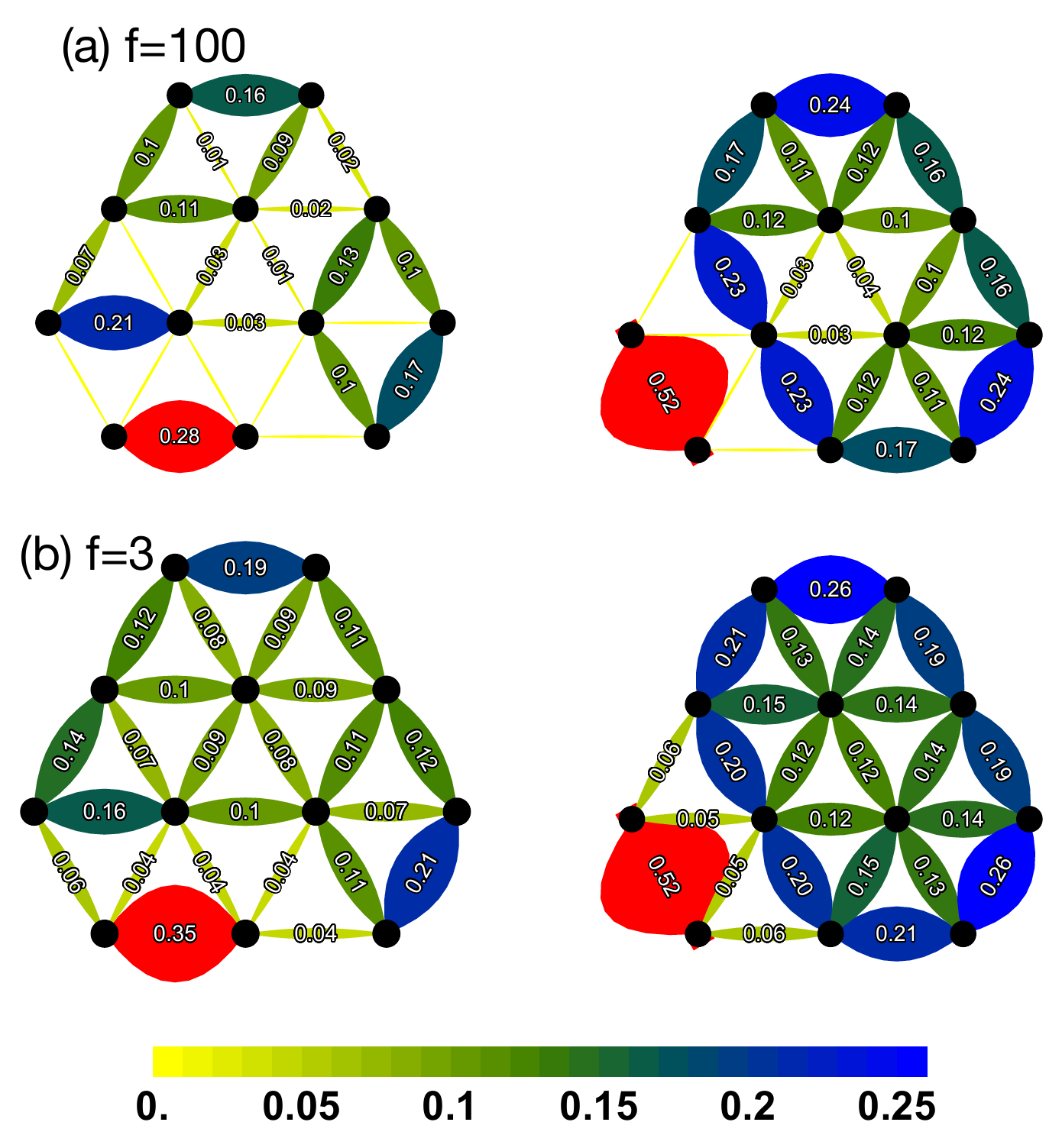}
\caption{(Color online) Dimer-dimer correlations $\langle \hA_{ij}\+\hA_{ij}\hA_{kl}\+\hA_{kl}\rangle$ on a triangular lattice for (a) $f=100$ and (b) $f=3$. The reference dimer $(i,j)$ is the fat red bond in the lower left corner, while $(k,l)$ is the bond located at the position of the line. The thickness of the lines is proportional to the strength of the correlation, which is also indicated by the color.
}
\label{fig:triangular lattice}
\end{figure}

Figure~\ref{fig:square lattice} shows the calculated correlations between a horizontal bond $(i,j)$ and a bond $(k,l)$ in the ground state on a square lattice for two different values of $f$, where the bond $(i,j)$ is fixed in the lower left corner, and the bond $(k,l)$ is varied. 
The fixed bond is colored red, while the other is colored based upon the strength of the correlations. A distinctive ``ladder'' pattern can be observed at $f=100$. This bond configuration is characteristic of the ``columnar state.'' Our large $f$ expansion in Sec.~\ref{largef} indeed confirms that the columnar state is expected to be the ground state at large $f$. In a finite size sample, the idealized columnar state is a quantum superposition of four symmetry related dimer crystals, as shown in Fig.~\ref{columnar}. Within this cartoon, and taking $f\to\infty$, all correlations
$C_{ij}^{kl}=\langle \hA_{ij}^\dagger\hA_{ij}\hA_{kl}^\dagger\hA_{kl}\rangle$ will be either 0 or 0.25, depending on the two bonds $(i,j)$ and $(k,l)$ are both found in the same configuration. The reference value should be $C^{ij}_{ij}=0.25$. The correlations in Fig.~\ref{fig:square lattice}(a) share this same pattern, but the contrast is somewhat weaker than in the idealized picture. Such ``quantum fluctuations'' are due to the fact that the quantum state has weight on configurations other than those given by this cartoon.

For $f=3$ [Fig.~\ref{fig:square lattice}b], the correlations are somewhat more ambiguous. The pattern includes short-range columnar order, but it is unclear if there is long-range order.

\begin{figure}[t]\centering
\includegraphics[width=1.\columnwidth]{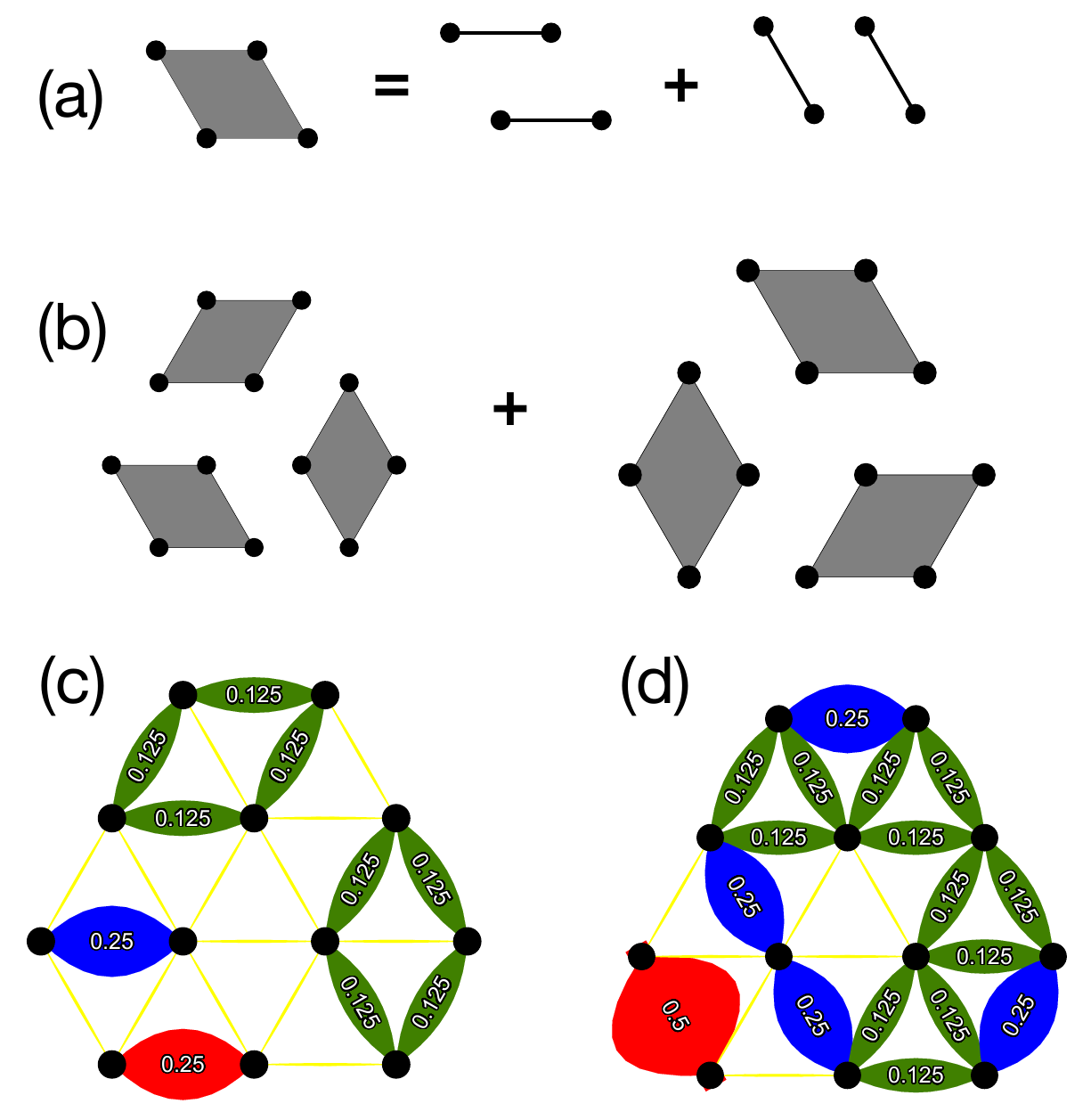}
\caption{\label{triexpected}
(Color online) Cartoons of bond patterns in $\sqrt{12}\times\sqrt{12}$ dimer crystal phase on triangular lattice. (a) Each shaded rhombus represents a quantum superposition of two bond patterns. (b) The $\sqrt{12}\times\sqrt{12}$ phase is idealized as a quantum superposition of two patterns of resonating bonds. This cartoon corresponds to a variational wave function from which one can calculate the bond correlations 
$C_{ij}^{kl}=\langle \hA_{ij}^\dagger\hA_{ij}\hA_{kl}^\dagger\hA_{kl}\rangle$. In (c) and (d), we take the lower left (fat red) bond as the reference $(i,j)$. }
\end{figure}

 Figure~\ref{fig:triangular lattice} shows the correlations on a triangular lattice for the same two values of $f$. For this lattice, we anticipate the ground state to be the $\sqrt{12}\times\sqrt{12}$ phase \cite{Ralko2005}, and so we take our system to have the shape of a unit cell in the $\sqrt{12}\times\sqrt{12}$ phase. 
 As illustrated in Fig.~\ref{triexpected}(b), the expected unit cell consists of 12 sites, which resonate between two plaquette configurations. Figure~\ref{triexpected}(a) further illustrates that in each plaquette the spins are expected to resonate between two different dimer configurations. The correlations corresponding to this ansatz are shown in Figs.~\ref{triexpected}(c) and ~\ref{triexpected}(d), for two different reference bonds. The thicker (blue) bonds in the idealized model have $C=0.25$, while thinner (green) bonds have $C=0.125$, as labeled. The interpretation is that when one expands out the superposition in Fig.~\ref{triexpected}(b), 1/4 of the terms will simultaneously have bonds at a given red and blue position, and 1/8 of the terms will simultaneously have bonds at a given red and green position. The red bond is thicker in Fig.~\ref{triexpected}(d) than~\ref{triexpected}(c), as it appears in more terms of the superposition.
 
 For large $f$, the pattern of bonds in Fig.~\ref{fig:triangular lattice} are nearly identical to what one expects from the $\sqrt{12}\times\sqrt{12}$ phase. The deviations are of comparable size to those seen in the square lattice.
The $f=3$ pattern shares some of the same symmetries, but one observes significant differences, which can be interpreted as spatial broadening. For example, in the lower image of Fig.~\ref{fig:triangular lattice}, the central three bonds have relatively substantial weight, while no such weight is found in the cartoon of the $\sqrt{12}\times\sqrt{12}$ phase. Similar discrepancies were seen in the numerical studies of the models in Ref.~\cite{Ralko2005}. Since our exact diagonalization approach is only able to capture a single unit cell, we cannot say anything about long-range order from this calculation.

We also use these numerical results to investigate the stability of our system against small perturbations, such as off-site dipole-dipole interactions and magnetic field noise. Using parameters appropriate for Lithium, we find that the amplitude for off-site dipole interactions to create an excitation is only $\sim 0.008\ {\rm Hz}$, which is small compared to the lowest energy plaquette-flip excitation $\sim 0.9J^2/U_0 \sim 90\ {\rm Hz}$. Our system is however sensitive to small perturbations in the magnetic field, which will break dimers to align spins in its direction. In order to remain in the ground state, the magnetic field in the experiment should be less than $30\mu$G.

\section{Large $f$ limit}\label{largef}
\subsection{Mapping onto the Rokhsar-Kivelson model}\label{rkm}
We now consider the large $f$ limit, and show that our model maps onto the classic Rokhsar-Kivelson dimer model.

For notational convenience it is useful to introduce a fictitious Hilbert space with orthonormal basis states labeled by the singlet coverings $|\tilde a\rangle$. This allows us to use the familiar language of bras and kets in manipulating $H_{ab}$. For example, we write $\tilde H = \sum_{ab} H_{ab} |\tilde a\rangle \langle \tilde b |$. An eigenstate $|\tilde \psi\rangle$ of $\tilde H$ can be mapped into the physical space via $|\psi\rangle = \hat P|\tilde \psi\rangle$, where $\hat P=\sum_a |a\rangle \langle \tilde a|$. All eigenvalues of $\tilde H$ are also eigenvalues of $H$, with the caveat that zero energy eigenvectors may be unphysical as quantum states. 

The diagonal elements of $\tilde H$, $H_{aa}=(-2 J^2/U_0) N_a$ count the number of nearest-neighbor singlets in $|a\rangle$. The low-energy space is then spanned by nearest-neighbor singlet coverings. 
We will elimate the other modes to derive an effective Hamiltonian which acts only in this low-energy space. The key point is that
the off-diagonal matrix elements are of order $(2f+1)^{-1}$, and are small in the limit $f\to \infty$. 

The leading order term in the effective Hamiltonian comes from the parts of $\hat H$ which directly take one between nearest-neighbor singlet coverings. For example, the $\hat A_{14}^\dagger \hat A_{14}$ or $\hat A_{25}^\dagger \hat A_{25}$ terms acting on the state $|a\rangle$ in Fig.~\ref{fig:ActionOnKets}. Terms of this form take two parallel vertical nearest-neighbor bonds, and replaces them with horizontal bonds (or vice versa), and can be represented as $\tilde H_0=-t\left(\ket{\tilde=}\bra{\tilde\shortparallel} + \ket{\tilde\shortparallel}\bra{\tilde=}\right)$ with $t=(2 J^2/U_0) (2/(2f+1))$. The factor of 2 comes from the fact that the same term is generated if one acts on either of the new bonds.

\begin{figure}
\includegraphics[width=\columnwidth]{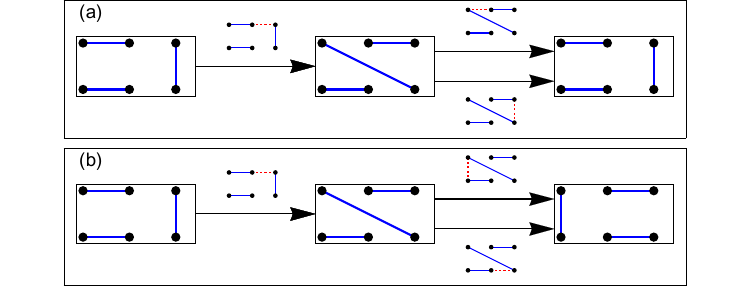}
\caption{\label{secorder} (Color online) Illustration of second order process which takes one out of the space of nearest-neighbor singlet coverings, then back. At second order, these fall into two classes: (a) cyclic process that return the system to original state and (b) six-site ring exchange processes. 
}
\end{figure}

Acting on any of the other nearest-neighbor bonds will introduce a long-range bond. For example, the $\hat A_{23}^\dagger \hat A_{23}$ term acting on the state $|a\rangle$ in Fig.~\ref{fig:ActionOnKets}, yields the state with long range bonds on the far right. There are $3N/2-2 (N_{=}+N_{\shortparallel})$ such terms, each of which will contribute to the effective Hamiltonian in second-order perturbation theory. Here $N_{=}$ and $N_{\shortparallel}$ count the number of plaquettes with two horizontal or vertical bonds. As illustrated in Fig.~\ref{secorder}(a), for each of these terms, there are two ways to return to the initial state. 
Up to an additive constant, one thus finds a contribution to the effective Hamiltonian of $\tilde H_{1a} = V \left(\ket{\tilde=}\bra{\tilde{=}} + \ket{\tilde\shortparallel}\bra{\tilde{\shortparallel}}\right)$ with $V= 2\times\lambda\times(2\lambda)/\epsilon= (8 J^2/U_0)/(2f+1)^2$. Here one factor of 2 comes from bond counting, one $\lambda=(2J^2/U_0)/(2f+1)$ is for the forward matrix element, the $2\lambda$ is for the backward matrix element, and $\epsilon=(2J^2/U_0)$ is the energy denominator.

There are also second order processes which, as illustrated in Fig.~\ref{secorder}(b), rotates a set of three bonds, and can be written as
$\hat H_{1b}= - t'\left(\ket{\widetilde{=\shortmid}}\bra{\widetilde{\shortmid=}} + \rm{h.c.}\right)$, with $t^\prime= (2\lambda)^2/\epsilon_0=(8J^2/U_0)/(2f+1)^2$. Here one factor of $2$ comes from the existence of two possible intermediate states (each of which can be reached in a single way). The second factor of $2$ comes from the two ways to reach the rotated configuration from each intermediate state.

This reasoning can be continued to generate terms involving longer and longer ring exchanges. Any term which appears at $m^{\rm th}$ order scales as $(2f+1)^{-m}$. Thus, unlike the spin-1/2 electronic case, larger ring exchange terms are strongly suppressed for large $f$.

This same argument goes through on any lattice. On a square, cubic, or triangular lattice, the effective Hamiltonian is of the form
\begin{align} \label{eqn: dimer model}
 \hH_{\rm QDM} = &\sum-t\left(\ket{\tilde{=}}\bra{\tilde{\shortparallel}} + \ket{\tilde{\shortparallel}}\bra{\tilde{=}}\right) + V\left(\ket{\tilde{=}}\bra{\tilde{=}} + \ket{\tilde{\shortparallel}}\bra{\tilde{\shortparallel}}\right)\nonumber \\& - t'\left(\ket{\widetilde{=\shortmid}}\bra{\widetilde{\shortmid=}} + \ket{\widetilde{\shortmid=}}\bra{\widetilde{=\shortmid}}\right)+\cdots
\end{align}
This defines the matrix elements $H_{ab}$. Table~\ref{table} lists the parameters $t, t'$, and $V$ for different lattice geometries. On the triangular lattice one interprets parallel nearest-neighbor bonds as those that are on opposite sides of a rhombus made from two triangular units.  The effective dimer model on the honeycomb and Kagome lattices have similar terms, but the smallest kinetic term involves three-bond loops, and therefore $t$ has an amplitude of $\mathcal{O}\left(2f+1\right)^{-2}$. Similarly, on those lattices the potential term $V\sim \mathcal{O} \left(2f+1\right)^{-4}$ penalizes parallel bonds on alternate sides of a hexagon, while $t^\prime \sim \mathcal{O} \left(2f+1\right)^{-4}$ involves a ring with bonds extending over two hexagons.

\begin{table}[t]
\centering
\begin{tabular}{ccccc}
\hline\hline
{\rm Lattice geometry} & $\frac{t}{J^2/U_0}$ & $\frac{V}{J^2/U_0}$ & $\frac{t'}{J^2/U_0}$ & Ground state\\
& & & & at large $f$ \\ \hline
{\rm Square lattice} & $\frac{4}{2f+1}$ & $\frac{8}{(2f+1)^2}$ & $\frac{8}{(2f+1)^2}$ & columnar$^a$\\
{\rm Triangular lattice} & $\frac{8(f+1)}{(2f+1)^2}$ & $\frac{4}{(2f+1)^2}$ & $\frac{8}{(2f+1)^2}$ & $\sqrt{12}\times\sqrt{12}$\\
{\rm Cubic lattice} & $\frac{4}{2f+1}$ & $\frac{8}{(2f+1)^2}$ & $\frac{8}{(2f+1)^2}$ & \\
{\rm Honeycomb lattice} & $\frac{12}{(2f+1)^2}$ & $\mathcal{O}\left(1/f^4\right)$ & $\mathcal{O}\left(1/f^4\right)$ & plaquette\\
{\rm Kagome lattice} & $\frac{12}{(2f+1)^2}$ & $\mathcal{O}\left(1/f^4\right)$ & $\mathcal{O}\left(1/f^4\right)$ & \\
\hline\hline
\end{tabular}
\caption{List of ring exchange amplitudes and bond interactions obtained from Eq.~\eqref{eqn: AdagA}, for different lattice geometries. $^a$There is some debate in the literature about the phases of dimer models on a square lattice \cite{Banerjee2014}.}
\label{table}
\end{table}

\subsection{Phases}\label{sec: phases}
The effective model for our system from Eq.~\eqref{eqn: dimer model} has a rich phase diagram, which has been well explored along $t'=0$ in a number of geometries \cite{Moessner2008, Sachdev1989, Moessner2001a, Moessner2001b, Moessner2001c, Moessner2002, Misguich2002, Ralko2005, Zeng1995, Leung1996, Syljuasen2006, Ralko2008, Banerjee2014}. For 2D bipartite lattices with $t'=0$, one finds only valence bond solid phases, except for the Rokhsar-Kivelson point $V=t$. On 3D and nonbipartite 2D lattices, dimer liquids may be found for nonvanishing ranges of $t/V$. The phase diagram at finite $t'$ is less explored \cite{Nakata2011}.

The valence bond solid phases described in the literature fall into four types: columnar, plaquette, mixed, and staggered. The columnar phase is built from vertical columns of horizontal parallel bonds, or vice versa. In the plaquette phase, dimer bonds resonate between different configurations inside a multi-site unit cell. For example, on a square lattice, the plaquette phase has a unit cell with four lattice sites; two parallel bonds resonate between horizontal and vertical configurations inside a plaquette. The plaquette phases on a triangular lattice have larger unit cells. The mixed phase is a hybrid between the columnar and plaquette phases, which is best described in terms of the symmetries it breaks \cite{Banerjee2014}. The staggered phase has no flippable plaquettes ($=$ or $\shortparallel$). The columnar phase is favored at large negative $V$, and the staggered phase at large positive $V$. 

As $f\to\infty$, the dominant coupling constant in the effective model, Eq.~(\ref{eqn: dimer model}), is $t$. On a square lattice,
this generally is believed to lead to a columnar phase, though there is some contention~\cite{Banerjee2014}. (Experiments may be able to resolve these issues.)
On the triangular lattice, as $f\rightarrow\infty$, we expect to see a plaquette phase, called the $\sqrt{12}\times\sqrt{12}$ phase, which has a $12$-site unit cell, and quantum resonances that extend throughout the cell \cite{Ralko2005}. 
Observing these resonances is part of Pauling and Anderson's vision of quantum resonances that manifest throughout a macroscopic system~\cite{Anderson1973}. The analysis in Sec.~\ref{results} confirms that at $f=100$ these orders appear to be present. The smaller $f$ data is more ambiguous, and could point towards a spin liquid or some other phase.

\subsection{Tuning the parameters in the large $f$ dimer model}\label{tune}

As presented, a given experimental realization yields a unique dimer model: Aside from the overall energy scale, all parameters are
determined by the spin $f$ and the lattice geometry. We can gain the ability to tune the parameters by modifying the detuning and coupling strength of the optical Feshbach resonance. For example, consider the case
$U_{F\neq0,2}\gg U_2>U_0\gg J$. Then the effective Hamiltonian at second order in the tunneling strength is
\begin{equation}
 \hH_{\rm eff} = -\frac{2J^2}{U_0} \sum_{\langle ij\rangle} \hA_{ij}^{00\dagger}\hA_{ij}^{00} - \frac{2J^2}{U_2} \sum_{\langle ij\rangle M} \hA_{ij}^{2M\dagger}\hA_{ij}^{2M},
\end{equation}
where $\hA_{ij}^{2M\dagger} = \sum_m C_{m,M-m}^{2M}\hb_{i,m}\hb_{j,M-m}$ creates a neighboring atom pair with total spin $F=2$ and azimuthal spin $M$. In Appendix~\ref{extended} we use our large-$f$ perturbation techniques to obtain a dimer model of the form of Eq.~\eqref{eqn: dimer model}. We find that to leading order $t$ and $t^\prime$ are independent of $U_2$, but $V$ depends on $U_2$.
By tuning $U_2/U_0$ via the Feshbach laser intensity and detuning, one can control the relative size of $V$.

\section{Detection}\label{sec: detection}
To probe the valence bond solid order and observe the resonating dimers in the plaquette and spin liquid phase, we propose measuring the dimer-dimer correlation function $\langle \hA_{ij}\+\hA_{ij}\nodag\hA_{kl}\+\hA_{kl}\nodag\rangle$. 
Similar correlation functions have been used to characterize order in quantum dimer models~\cite{Misguich2002, Ralko2005, Zeng1995, Leung1996}. 
We provide an experimental protocol to image these correlations.
Furthermore, in Sec.~\ref{results} we numerically calculated these correlations in our system for both large and moderate values of $f$.

To image the dimer bonds we propose shining a weak near-resonant photoassociation laser on the system, tuned near a molecular state with angular momenta $L=1$ and $S=0$. In our system, when virtual hopping brings two atoms forming an $S=0, L=0$ dimer onto the same lattice site, the near-resonant light drives these atoms into the molecular state. The excited molecule has a short lifetime and so those atoms are lost from the trap.

After driving this photoassociation, one would
 use a quantum gas microscope to image the location of all remaining atoms \cite{Bakr2009, Jones2006}. All adjacent pairs of empty sites in the image were likely occupied by atoms entangled in dimers. In this way, a fraction of the dimers in the system can be imaged. Quantitative dimer-dimer correlations can be extracted by analyzing data from multiple realizations of this imaging process, and can be used to identify the phase.
 Similar techniques have been used in the past to probe atomic correlations~\cite{partridge2005molecular}.

One formal way to model this process is to take $U_0\to U_0+i\Gamma/2$, where $\Gamma$ quantifies the photoassociation rate. We thus see that the Hamiltonian in Eq.~(\ref{eqn: AdagA}) gains an imaginary term which removes a pair of neighboring particles. The probability that (after a fixed time) atoms at neighboring sites $i$ and $j$ are missing will be proportional to $\langle \hA_{ij}\+\hA_{ij}\nodag\rangle$. The probability that there are also missing atoms at neighboring sites $k$ and $\ell$ will then be proportional to $\langle \hA_{ij}\+\hA_{ij}\nodag\hA_{kl}\+\hA_{kl}\nodag\rangle$.

We emphasize that the ability to directly image the valence-bond correlations is one of the greatest strengths of using cold atoms to explore dimer models. This imaging will allow unambiguous identification of the various valence-bond ordered phases. Spin liquid phases will be characterized by the absence of long-range valence bond order. 
The experimental systems are much larger than those we can model numerically. 

\section{Summary}
In summary, we propose experimental protocols to produce quantum dimer models and detect both static and resonating patterns of dimer configurations. In particular, we show that appropriately tuned off-resonant photoassociation light modifies the interactions in a gas of cold atoms, yielding a low-energy Hilbert space spanned by short-range dimers. By expanding in powers of $(2f+1)^{-1}$ we develop an effective dimer model Hamiltonian, and discuss its phase diagram. We find that a number of valence bond solid and plaquette phases are readily produced, and suggest techniques which are suited to searching for even more exotic states such as topological spin liquids. We demonstrate that by combining photoassociation with quantum gas microscopy one can directly detect the dimers and the dimer-dimer correlations, thereby probing the defining features of these phases. We numerically calculate the dimer correlations, finding that on triangular lattices one will be able to image an intricate pattern of resonating bonds, extending over a 12-site unit cell.

Quantum dimer models have been highly influential in developing an understanding of how geometric constraints lead to new emergent physics~\cite{Moessner2002,Moessner2008}, and they have been used as a theoretical foundation for attempting to understand phenomena ranging from high temperature superconductivity to exotic antiferromagnets~\cite{Savary2017}. A direct experimental realization of dimer models is key to validating and refining these ideas. 

\section*{Acknowledgment}
This material is based upon work supported by the NSF Grants No. PHY-1508300, No. PHY-1806357, and No. PHY11-25915. This work was supported in part by the Data Analysis and Visualization Cyberinfrastructure funded by NSF under Grant No. OCI-0959097. B.S. acknowledges useful discussions with Randall Hulet, Anna Marchant and Jacob Fry. B.S. and T.C.R. contributed equally to this work.

\appendix
\section{Deriving the effective interaction due to an optical Feshbach resonance}\label{sec: deriving the effective interaction}
Here we derive the effective interaction induced by our proposed optical Feshbach resonance. We closely follow the argument in our previous work, Ref.~\cite{Sundar2016}. A laser is tuned near a transition to a molecule state with well-defined electronic spin $S=0$, and well-defined electronic angular momentum $L=1$.
Keeping only the relevant degrees of freedom, and neglecting any coupling to the nuclear degrees of freedom,
we model the photoassociation as
\begin{align}
\hat{H}_{\rm Fesh} =& \sum_{m,m'} \left(E+i\frac{\Gamma}{2}\right)\ket{\rm mol}_{mm'}\bra{\rm mol}_{mm'}\nonumber \\&+ \Omega\left(e^{-i\omega t}\ket{\rm mol}_{mm'}
\bra{\rm at}_{mm^\prime}+ {\rm h.c}\right),
\end{align}
where the electronic singlet state is
\begin{equation}\label{sing}
\ket{\rm at}_{mm^\prime}=
\frac{\ket{\ua m}\otimes\ket{\da m'}-\ket{\da m}\otimes\ket{\ua m'}}{\sqrt{2}} 
\end{equation}
Here, $\uparrow/\downarrow$ represent the spin projection $s_z$ of the spin-1/2 electrons, while $m,m^\prime$ are the spin projections of the nuclei.
Due to hyperfine interactions, $\ket{\rm at}_{mm^\prime}$ is not an eigenstate of the atomic Hamiltonian. 

The energy of the molecule is $E$, and we have included an imaginary part, $\Gamma$ to model its finite lifetime. In principle, the molecular energy should have some dependence on the nuclear spin projections, but these will play no role as long as the detuning of the laser is large compared to the hyperfine splitting. Thus, we ignore them. We further assume that the atoms are both in the same spatial mode of a single site of the optical lattice, and therefore drop spatial indices. The coupling $\Omega$ will depend on the shape of this mode. The laser frequency $\omega=E+\delta$ is detuned from the atom-molecule transition by $\delta$. When the detuning $\delta$ is large compared to $\Omega$, we use second-order perturbation theory to eliminate the molecule and obtain in a rotating frame
\begin{equation}
\hat{H'}_{\rm Fesh} = \frac{\Omega^2}{\delta+i\Gamma/2}\sum_{mm'} \ket{\rm at}_{mm^\prime}\bra{\rm at}_{mm^\prime}
\end{equation}
If the incoming and outgoing atoms are restricted to being in a single hyperfine manifold ($f=i\pm 1/2$), then symmetry implies that this expression can be replaced by
\begin{eqnarray}
\hat{H'}_{\rm Fesh}^f&=& \hat P_f \hat{H'}_{\rm Fesh} \hat P_f\\
&=&\frac{\Omega^2}{\delta+i\Gamma/2}\sum_{FM}\alpha_F^f\ket{F,M}_f \bra{F,M}_f,
\end{eqnarray}
where $|F,M\rangle_f$ is the two-particle state with total hyperfine spin $F$ and total spin projection $M$, built from two particles in the manifold with hyperfine spin $f$. The operator $P_f$ projects into the space where each atom has spin projection $f$. The $SU(2)$ symmetry implies that the coefficients $\alpha_F^f$ do not depend on $M$. In the main text we do not explicitly write the $f$ labels.
Equating these expressions for $\hat{H}'_{\rm Fesh}$ gives
\begin{eqnarray}\label{msum}
\alpha_F^f &=&\sum_{m} 
\left| \langle F, M|_f \ket{\rm at}_{m(M-m)}\right|^2,
\end{eqnarray}
where the state $ \ket{\rm at}_{m(M-m)}$ is given by Eq.~(\ref{sing}) with $m^\prime=M-m$. Alternatively, this can be written as the square overlap of two states: the first is formed by combining $i_1$ and $s_1$ into $f_1$, $i_2$ and $s_2$ into $f_2$, then $f_1$ and $f_2$ into $F$. The second is formed by combining $s_1$ and $s_2$ into $S$, $i_1$ and $i_2$ into $I$, then $S$ and $I$ into $F$. The nine angular momenta $s_1,s_2,i_1,i_2,f_1,f_2,S,I,F$ can be combined into a Wigner 9-j symbol \cite{NIST:DLMF}. The most natural notation for this construction involves
recursively noting how each angular momentum is constructed. For example $f_1(i_1 s_1)$ indicates that $f_1$ is built from $i_1$ and $s_1$. In this notation,
\begin{equation}\label{ninej}
\alpha_F^f = |\bra{F(f_1 (i_1 s_1) f_2(i_2 s_2))}{F (I (i_1 i_2)S(s_1 s_2))}\rangle|^2.
\end{equation}
Here we have a particularly simple case where $s_1=1/2,s_2=1/2,i_1=i,i_2=i,f_1=f,f_2=f,S=0,I=F$.  A third representation of the coefficient is the expectation value
\begin{equation}\label{expect}
\alpha_F^f= \langle F,M| P_{S=0} |F,M\rangle,
\end{equation}
where $P_{S=0}=(1/4)-{\bf S_1\cdot S_2}$ is the projector into the space where ${\bf S}={\bf S_1+ S_2}=0$.

There are several ways to evaluate $\alpha_F$. The simplest is to note that the condition $S=0$ reduces Eq.~(\ref{ninej}) to a 6-j symbol---which is tabulated in Ref.~\cite{NIST:DLMF} for the case $s_1=s_2=1/2$. The second is to directly evaluate Eq.~\eqref{msum}. Presumably there is also an approach based upon Eq.~\eqref{expect}. In the remainder of this section we outline the second method, based on Eq.~\eqref{msum}.
 
We first note that since the result is independent of $M$, we can set $M=0$. 
 We then find a common basis for each set of states, using the Clebsch-Gordon coefficients, defined by
 \begin{eqnarray}
|f m_f\rangle &=& \sum_{m_s+m_i=m_f} C^{i+s\to f}_{m_i,m_s} |s m_s,i m_i\rangle,\\
|F M\rangle &=& \sum_{m_1+m_2=M} C^{f+f\to F}_{m_1,m_2} |f m_1, f m_2\rangle.\label{fm}
 \end{eqnarray}
Using tabulated expressions for $S=1/2$, we can invert this relationship to arrive at
\begin{align}
\ket{s\sigma, im} &= \sigma \sqrt{\frac{f+1/2-\sigma m}{2i+1}}\ket{f=i-\frac{1}{2},m_f=m+\frac{\sigma}{2}}\nonumber\\& + \sqrt{\frac{f+1/2+\sigma m}{2i+1}}\ket{f=i+\frac{1}{2},m_f=m+\frac{\sigma}{2}},
\label{eqn: clebschgordan}
\end{align}
where $\sigma=+1(-1)$ corresponds to $\ua(\da)$, and as before $m$ is the nuclear spin projection. Substituting this result into Eq.~(\ref{sing}) and combining it with Eq.~(\ref{fm}) yields
\begin{eqnarray}\label{eqn: alphaF}
\alpha_F &=&\sum_m \left(
\lambda_{-m} C^{f+f\to F}_{m-1/2,-m+1/2}
 -\lambda_m C^{f+f\to F}_{m+1/2,-m-1/2}\right)^2,
 \nonumber\\
\lambda_m&=& \frac{f+1/2+2m(f-i)}{\sqrt{2}(2i+1)}.
\end{eqnarray}
We then derive a series of sum rules: First, we express $A=\sum_n \left(C^{f+f\to F}_{n,-n}\right)^2$, as
$A= \sum_n \langle F, 0| f, n;f -n\rangle \langle f, n;f -n | F,0\rangle$. This sum contains a resolution of the identity in the sector with $M=0$, and hence $A=\langle F, 0| F,0\rangle=1$. 
Second, by the same reasoning $B=\sum_n n^2 \left(C^{f+f\to F}_{n,-n}\right)^2$ $= \sum_n \langle F, 0| f, n;f ,{-n}\rangle n^2\langle f, n;f,-n | F,0\rangle$ which can be identified as the expectation value
$B=-\langle F, 0| \hat f_1^z \hat f_2^z |F,0\rangle$. Finally, $C=\sum_n(f-n)(f+n+1)\left(C^{f+f\to F}_{n,-n}\right)^2$ $=\langle F, 0| \hat f_1^+ \hat f_2^- |F,0\rangle$. Using the symmetry between the two spins, we can simplify this to $C=\langle F, 0| \hat {\vec f}_1 \cdot \hat {\vec f}_2-\hat f_1^z \hat f_2^z |F,0\rangle$. The resulting three identities are
\begin{align}
 \sum_n \left(C^{f+f\to F}_{n,-n}\right)^2 &= 1,\\\nonumber
 \sum_n n^2 \left(C^{f+f\to F}_{n,-n}\right)^2 &= -\langle F, 0| \hat f_1^z \hat f_2^z |F,0\rangle,\\\nonumber
  \sum_n (f-n)(f+n+1)& \left(C^{f+f\to F}_{n,-n}\right)^2 \\\nonumber
  &= \langle F, 0| \hat {\vec f}_1 \cdot \hat {\vec f}_2-\hat f_1^z \hat f_2^z |F,0\rangle.
\end{align}
These sum rules, plus the expression $\langle F,0| \hat F^2 |F,0\rangle= F (F+1)=2 f(f+1) + 2 \langle F,0| \hat {\vec f}_1 \cdot \hat {\vec f}_2 |F,0\rangle$, allow us to write Eq.~(\ref{eqn: alphaF}) as
\begin{equation}
\alpha_F^f = \frac{(2i+1)(2f+1)-F(F+1)}{2(2i+1)^2}.
\end{equation}
For our purposes, the most important feature of this expression is that it is monotonic in $F$.

\section{Construction of an orthogonal basis} \label{sec: orthogonal basis}

\begin{figure}
\includegraphics[scale=0.26]{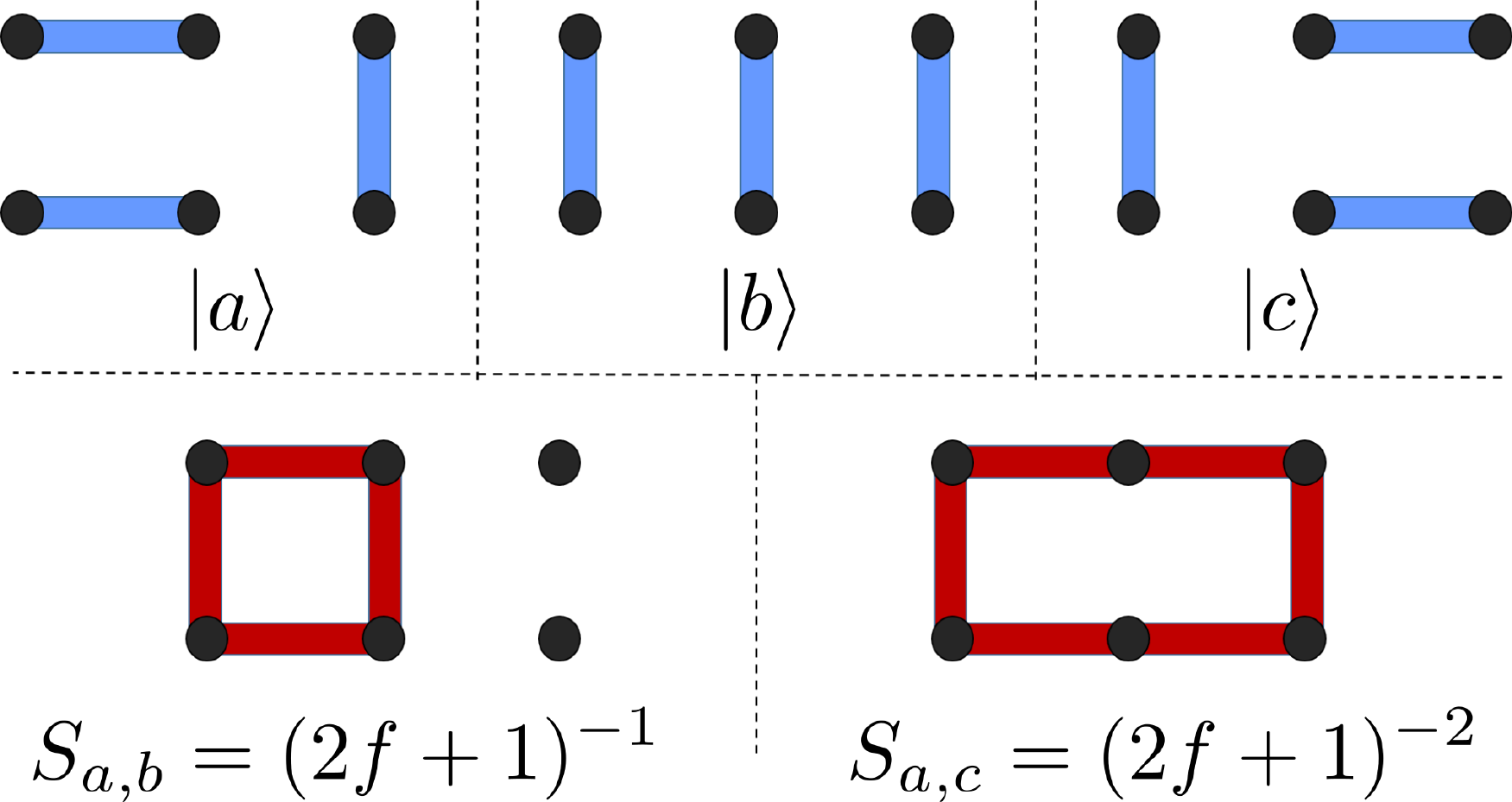}
\caption{\label{fig:transition_graph} (Color online) Examples of transition graphs between nonorthogonal singlet coverings, constructed graphically following Ref. \cite{Rokhsar1988}. The magnitude of the overlap is given by Eq.~\eqref{eqn: overlap} and is shown in the figure for the two cases. The overlap $S_{ab}$ comes from a single four-site loop, and it represents the largest possible overlap in magnitude. The overlap $S_{ac}$ comes from a single six-site loop, and it is down in magnitude by a factor of $(2f+1)^{-1}$. In the large-f limit, all singlet coverings become orthogonal as the overlaps approach zero.
}
\end{figure}
\begin{figure*}[t]
\includegraphics[scale=0.54]{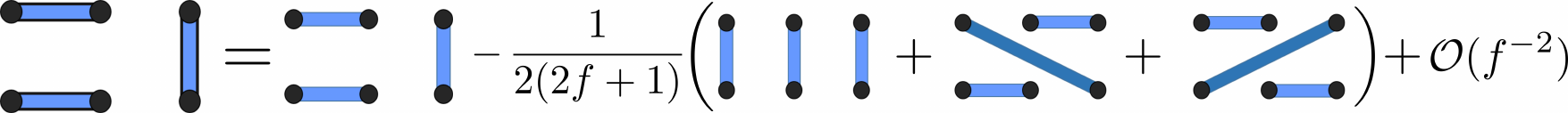}
\caption{\label{fig:OrthogonalState} (Color online) Pictorial representation of an orthogonal dimer state constructed from nonorthogonal singlet coverings, as expressed in Eq.~\eqref{eqn: dimer_basis}. A dimer state $\ket{\bar{a}}$ has an associated $\mathcal{O}(1)$ singlet covering $\ket{a}$, which is used to label the state. At $\mathcal{O}(f^{-1})$ and higher, it contains contributions from all coverings $\ket{b}$ which differ from $\ket{a}$ by a four-site loop in their transition graph, including those which lie outside the nearest-neighbor-only Hilbert space. In the $f \rightarrow \infty$ limit we find that the singlet coverings become orthogonal, such that $\ket{\bar{a}} = \ket{a}$
}
\end{figure*}

In the main text we work with a nonorthogonal basis, while the more traditional approach involves orthogonalizing the basis states, as originally developed for spin-$1/2$ systems in Ref. \cite{Rokhsar1988}. Here we follow a similar procedure to construct the orthogonal basis of dimer states for cold atoms for large spin $f$, perturbatively in $(2f+1)^{-1}$. The singlet coverings contain both short- and long-ranged singlet bonds. 

We first note that the singlet operators commute with one-another unless they share a site. Hence the overlap between two singlet coverings $S_{ab}=\langle a|b\rangle$ factors into expectation values of loops: sets of connected sites $\{i_1,i_2,\cdots i_L\}$. We can always label these loops so that $(i_{2j+1},i_{2j+2})\in a$ and $(i_{2j+2},i_{2j+3})\in b$ for $j=0,1,\cdots L/2-1$. We will also have $(i_L,i_1)\in b$. The contribution to $S_{ab}$ from such a loop is
\begin{equation}\label{loopcont}
S_{ab}^{\{i_1,i_2,\cdots i_L\}}=\langle A_{i_1i_2}A_{i_3,i_4}\cdots A_{i_{L-1}i_L} A^\dagger_{i_2,i_3} A_{i_4,i_5}^\dagger\cdots A_{i_L,i_1}^\dagger \rangle
\end{equation}
where the expectation value is in the vacuum state with no particles. The full $S_{ab}$ is the product of the contribution from all such loops.
To graphically generate this set of loops, one simply takes the set of all bonds which are in only one of $a$ and $b$, but not the other. Figure~\ref{fig:transition_graph} shows two examples of such a graphical construction of the overlap matrix elements for a six-site system.

Using the expression for the Clebsch-Gordan coefficients,
and assuming $i\neq j$, Eq.~(\ref{eqn: Aij}) becomes
\begin{equation}\label{exp}
A_{ij}=\frac{1}{\sqrt{2f+1}}\sum_m (-1)^{f-m} b_{im} b_{j-m}
\end{equation}
We substitute Eq.~(\ref{exp}) into Eq.~(\ref{loopcont}), and use Wick's theorem to evaluate the vacuum expectation value. There is only one nonzero contraction, as there is only one creation operator and only one annihilation operator acting on each site.  Once the $m$ of a single site is set, all others are fixed.  There are $2f+1$ choices for $m$, and each term contributes equally. Hence
$S_{ab}^{\{i_1,i_2,\cdots i_L\}}=(2f+1)^{1-L/2}$. 

The full expression for $S_{ab}$ is just the product of the contribution from each loop, and hence
\begin{equation}
S_{ab} = (2f+1)^{N_\text{loops}}\left(\frac{1}{\sqrt{2f+1}}\right)^{L_\text{loops}},
\label{eqn: overlap}
\end{equation}
where $N_\text{loops}$ is the total number of closed loops formed by the dimers not common to $\ket{a}$ and $\ket{b}$, while $L_\text{loops}$ is the total number of sites involved in all loops. 

For large $f$, we expand $S_{ab}$ in powers of $(2f+1)^{-1}$ to obtain
\begin{equation}
 S_{ab} = \delta_{ab} + \frac{\Box_{ab}}{2f+1} + \frac{\Box_{ab}^{(2)}}{(2f+1)^2} + \mathcal{O}(f^{-3}).
 \label{eqn: Sab}
\end{equation}
Here, $\Box_{ab}=1$ if $\ket{a}$ and $\ket{b}$ differ by a four-site loop in their transition graph, and is zero otherwise. The sites making up the loops do not need to be nearest neighbors. The symbol $\Box^{(2)}_{ab} =1 $ if $\ket{a}$ and $\ket{b}$ differ by either a single six-site loop, or two distinct four-site loops in their transition graph, and is zero otherwise. 
 
We can now construct orthogonal ``dimer states'' via
\begin{equation}
\label{eqn: dimer_basis}
\ket{\bar{a}} = \sum_{b} \left(\sqrt{S^{-1}}\right)_{a,b} \ket{b}.
\end{equation}
 The expansion in Eq.~(\ref{eqn: Sab}) formally leads to
\begin{align}
 \ket{\bar{a}} = &\ket{a} - \sum_b\left(\frac{\Box_{ab}}{2(2f+1)} + \frac{\Box^{(2)}_{ab}}{2(2f+1)^2}\right.\nonumber\\
 & \left.- \frac{3}{8(2f+1)^2}\sum_c \Box_{ac}\Box_{cb}+\cdots\right)\ket{b}. \label{eqn: orthogonal state}
\end{align}
Figure~\ref{fig:OrthogonalState} shows this construction for a small system of six sites. Although this expansion is commonly used in the literature \cite{Rokhsar1988}, it is at best formal. Intuitively one expects that for a given $f$, any given orthogonal dimer state $|\bar a\rangle$ will differ from the singlet covering $|a\rangle$ by a finite density of loops.  This intuition is reflected in the fact that subsequent terms in Eq.~(\ref{eqn: orthogonal state}) contain ever higher factors of the volume of space, and the limit $f\to\infty$ does not commute with the thermodynamic limit. For example, the diagonal element of third term in parentheses, $\sum_c \Box_{ac}\Box_{ca}$, is proportional to the total number of four-site loops that can be constructed, and scales as $N_b^2$, where $N_b$ is the number of bonds in $a$.  

Despite its formal nature, we note that one can use this expansion to derive an effective dimer model Hamiltonian. We omit the details as it is a lengthy argument, and the result is the same as we found in Sec.~\ref{sec: model derivation}.

\section{Beyond Singlets}
\label{extended}
In this appendix we analyze the case of Eq.~(\ref{heff}) to the case when two of the terms are significant, namely,
\begin{equation}\label{heff2}
\hat H_{\rm eff} = \sum_{\langle i j \rangle}-\frac{2 J^2}{U_0}\hA_{ij}^{00\dagger}\hA_{ij}^{00}
-\sum_{M=-2}^2\frac{2 J^2}{U_2}\hA_{ij}^{2M\dagger}\hA_{ij}^{2M},
\end{equation}
where as in the main text, $\hA_{ij}^{FM\dagger}$ creates a pair on sites $i$ and $j$ with total spin $F$ and spin projection $M$.
We will assume that $U_0\ll U_2$, so that the first term is large compared to the second, but that these two terms are large compared to all others.

The space of singlet coverings is not closed under Eq.~(\ref{heff2}), and we must enlarge our Hilbert space to include coverings with both spin-2 and spin-0 dimers---the former of which carry a quantum number $M$. For example, given two sites we have a six-dimensional Hilbert space, spanned by the singlet dimer, and the five spin-2 dimers. Given four sites, our Hilbert space is spanned by $3\times 6^2$ states---corresponding to the three different ways to pair up the four sites, and the six different flavors of each dimer. As in the purely singlet case, these states are not orthogonal. That is, the state $|\raisebox{-1.6mm}{\includegraphics[width=1.5em]{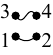}}\rangle=\hA_{12}^{FM\dagger}\hA_{34}^{F^\prime M^\prime \dagger}|{\rm vac}\rangle$ is not orthogonal to the state
$|\raisebox{-1.6mm}{\includegraphics[width=1.5em]{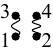}}\rangle=\hA_{14}^{F^{\prime\prime}M^{\prime\prime}\dagger}\hA_{23}^{F^{\prime\prime\prime} M^{\prime\prime\prime} \dagger}|{\rm vac}\rangle$, regardless of angular momenta and projections, which are denoted by the different styles of lines joining the sites. Different flavor bonds on the same sites, however, are orthogonal: $|\raisebox{-0.2mm}{\includegraphics[width=1.5em]{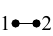}}\rangle=\hA_{12}^{00\dagger}|{\rm vac}\rangle$ is orthogonal to $|\raisebox{-0.2mm}{\includegraphics[width=1.5em]{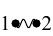}}\rangle=\hA_{12}^{2M\dagger}|{\rm vac}\rangle$.

The low-energy space is spanned by nearest-neighbor singlets. As described in the main text, acting on states of this form with $\hA_{ij}^{00\dagger}\hA_{ij}^{00}$ can either move us in this space, or generate longer-range singlets. We need to calculate how operators of the form $\hA_{ij}^{2M\dagger}\hA_{ij}^{2M}$ act on these states. Let $a$ describe the singlet covering. If $(i,j)\in a$, then $\hA_{ij}^{2M\dagger}\hA_{ij}^{2M}|a\rangle=0$. If $(i,j)\not\in a$ then the action of $\hA_{ij}^{2M\dagger}\hA_{ij}^{2M}$ will involve the sites $i,j$ and their partners $k,l$. No other bonds matter, so we consider the action on $|\raisebox{-1.6mm}{\includegraphics[width=1.5em]{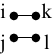}}\rangle=\hA_{ik}^{00\dagger}\hA_{jl}^{00\dagger}|{\rm vac}\rangle$. The notation does not imply any spatial relationship between the sites---just that they are connected. We then calculate $|\psi_M\rangle= \hA_{ij}^{2M\dagger}\hA_{ij}^{2M} |\raisebox{-1.6mm}{\includegraphics[width=1.5em]{aijklstraight.pdf}}\rangle$ as
\begin{eqnarray}
|\psi_M\rangle&=& \hA_{ij}^{2M\dagger}\hA_{ij}^{2M}\hA_{ik}^{00\dagger}\hA_{jl}^{00\dagger}|{\rm vac}\rangle\\
&=& \hA_{ij}^{2M\dagger} \sum_{mnp} \frac{C_{m,M-m}^{2M} }{2f+1}(-1)^{n+p}\\\nonumber
&&\qquad \times b_{i,m}b_{j,M-m} b_{in}^\dagger b_{k,\text{-}n}^\dagger b_{jp}^\dagger b_{l, \text{-}p}^\dagger |{\rm vac}\rangle\\
&=& \hA_{ij}^{2M\dagger} \sum_m \frac{C_{m,M-m}^{2M} }{2f+1} b_{k,\text{-}m}^\dagger b_{\ell,m-M}^\dagger |{\rm vac}\rangle\\
&=& \frac{1}{2f+1} \hA_{ij}^{2M\dagger}  \hA_{kl}^{2,\text{-}M\dagger} |{\rm vac}\rangle\\
&\equiv&
\frac{1}{2f+1} 
|\raisebox{-1.6mm}{\includegraphics[width=1.5em]{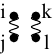}}_{M}\rangle,
\end{eqnarray}
where we have used $C^{00}_{n,-n}=(2 f+1)^{-1/2} (-1)^{f-n}$. The Hamiltonian will always generate a superposition of all different $M$, $|\psi\rangle=\sum_M |\psi_M\rangle.$ Returning to the space containing only singlets requires acting with either
$\hA_{ik}^{00\dagger}\hA_{ik}^{00}$ or $\hA_{jl}^{00\dagger}\hA_{jl}^{00}$. We therefore calculate
$|\phi\rangle=$ $\hA_{ik}^{00\dagger}\hA_{ik}^{00} |\psi\rangle$
$=\hA_{jl}^{00\dagger}\hA_{jl}^{00} |\psi\rangle$
as
\begin{eqnarray}
|\phi\rangle
&=& \frac{\hA_{ik}^{00\dagger}}{2f+1} \sum_{Mmnp} \frac{C_{m,M-m}^{2M}C_{-n,n-M}^{2,-M}}{\sqrt{2f+1}}(-1)^{f-p}\\\nonumber
&&\qquad \times b_{ip}b_{k{\text{-}p}}b_{im}^\dagger b_{j,M{-m}}^\dagger b_{k,-n}^\dagger b_{l,n-M}^\dagger |{\rm vac}\rangle\\
&=&\frac{\hA_{ik}^{00\dagger}}{2f+1} \sum_{Mm}\left(C_{m,M-m}^{2M}\right)^2  \frac{(-1)^{f-m}}{\sqrt{2f+1}}b_{j,m}^\dagger b_{l,-m}^\dagger|{\rm vac}\rangle.\nonumber
\end{eqnarray}
Below we show that $\sum_M \left(C_{m,M-m}^{2M}\right)^2 =5/(2f+1)$, which then gives
\begin{equation}
|\phi\rangle = \frac{5}{(2f+1)^2} \hA_{ik}^{00\dagger}\hA_{jl}^{00\dagger}|{\rm vac}\rangle.
\end{equation} 

This last Clebsch-Gordan identity is a special case of the more general result $V_{Ffm}=\sum_M (C^{FM}_{m,M-M})^2=(2F+1)/(2f+1)$, which is proven by writing
\begin{eqnarray}
V_{Ffm}&=&\sum_M \langle fm;fM-m|F,M\rangle^2 \\
&=& \sum_{Mn} \langle fm;fn|F,M\rangle^2 \\
&=& \langle fm| \hat X |fm \rangle,
\end{eqnarray}
where $\hat X={\rm Tr}_2 \sum_{M} |F,M\rangle\langle F,M|$ is the trace over the second spin of the projector into the space of fixed $F$. This operator clearly transforms as a singlet under rotation, and hence $V_{Ffm}$ must be independent of $m$. We therefore sum over $m$, and divide by $2f+1$ to find 
\begin{eqnarray}
V_{Ffm}&=&\frac{1}{2f+1}\sum_{Mnm^\prime}\langle fm^\prime;fn|F,M\rangle^2 \\
&=& \frac{1}{2f+1} \sum_M {\rm Tr} |F,M\rangle\langle F,M|\\
&=&\frac{2F+1}{2f+1}.
\end{eqnarray}

Having established the action of the various terms in the Hamiltonian on dimer coverings, we perturbatively eliminate the coverings, which either contain longer range singlet bonds, or any $F=2$ bonds. This process gives us an effective model which only involves nearest-neighbor singlets. The contributions from long-range singlets are identical to those derived in Sec.~\ref{sec: model derivation}. Below we show that the leading contributions from the $F=2$ bonds renormalize $V$, while leaving $t$ and $t^\prime$ unchanged.

Let $a$ be a nearest-neighbor singlet covering, and consider nearest neighbors $i$ and $j$ such that $a$ does not contain the bond connecting them: {\em i.e.}, $(i,j)\not\in a$. We will separately consider the case where the partners of $i$ and $j$ are also nearest neighbors, and the case where they are not. The first circumstance corresponds to parallel bonds. In that case, acting with $\hA_{ij}^{2M\dagger}\hA_{ij}^{2M\dagger}$ yields a state with two-fewer nearest-neighbor singlet bonds, but two extra nearest-neighbor $F=2$ bonds, and hence an excited state with energy $4 J^2/U_0 - 4 J^2/U_2$. In the second case, one instead finds an intermediate state with two-fewer nearest-neighbor singlets, but only one extra nearest-neighbor $f=2$ bonds. The other $F=2$ bond is long-ranged. Thus the second order process in which one returns to the initial state will have different coefficients for parallel and nonparallel bonds, hence shifting $V$. There will also be an unimportant constant energy shift to all states. In particular, the change in $V$ will be
\begin{equation}
\delta V=-(2\times \lambda^\prime)\times\left(\frac{1}{\epsilon}-\frac{1}{\epsilon^\prime}\right)\times(2\times \bar\lambda),
\end{equation}
where $\lambda^\prime=(2J^2/U_2) (2f+1)^{-1}$ is the forward amplitude, $\bar\lambda= 5(2J^2/U_0)(2f+1)^{-2}$ is the backward amplitude. The factors of $2$ account for the multiplicity of processes: There are two ways to produce a given intermediate state, and two ways back. The energy denominators 
$\epsilon=4J^2/U_0-4J^2/U_2$ and $\epsilon^\prime= (4J^2/U_0-2J^2/U_2)$ are the energy denominators associated with parallel and nonparallel bonds. We note that this shift is of $\mathcal{O}(f^{-3})$ and so the effect of sub-dominant scattering channels remain negligible in the large $f$ limit, furthering the validity of our description in terms of quantum dimers.

\bibliography{BibFile}

\end{document}